\documentclass[letterpaper]{emulateapj}

\usepackage{amsmath}
\usepackage{amssymb}
\usepackage{xspace}
\usepackage{color}
\usepackage{ifthen}
\usepackage{xspace}

\newcommand{\Msun}{{\ensuremath{\mathrm{M}_{\odot}}}\xspace}

\slugcomment{Draft for ApJ, \today}

\usepackage{ulem}
\begin{document}

\title{On the Sensitivity of Massive Star Nucleosynthesis and Evolution to Solar Abundances and to Uncertainties in Helium Burning Reaction Rates}

\author{Clarisse~Tur}
\affil{National Superconducting Cyclotron Laboratory,\\
Michigan State University, 1 Cyclotron Laboratory, East Lansing, MI 48824-1321\\
Joint Institute for Nuclear Astrophysics} 
\email{tur@nscl.msu.edu}

\author{Alexander~Heger}
\affil{Theoretical Astrophysics Group, MS B227, Los Alamos
National Laboratory, Los Alamos, NM 87545;\\
Department of Astronomy and Astrophysics, University of
California, Santa Cruz, CA 95064 \\
Joint Institute for Nuclear Astrophysics} 
\email{alex@ucolick.org}

\author{Sam~M.~Austin}
\affil{National Superconducting Cyclotron Laboratory,\\
Michigan State University, 1 Cyclotron Laboratory, East Lansing, MI 48824-1321\\
Joint Institute for Nuclear Astrophysics\phantom{.}}
\email{austin@nscl.msu.edu}

\begin{abstract}

We explore the dependence of pre-supernova evolution and supernova
nucleosynthesis yields on the uncertainties in helium
burning reaction rates. Using the revised solar abundances of
\cite{lod03} for the initial stellar composition, instead of those
of \cite{and89}, changes the supernova yields and limits the constraints
that those yields place on the $^{12}C(\alpha,\gamma)^{16}O$ reaction
rate. The production factors of medium-weight elements (A = 16-40) 
were found to be in reasonable agreement with observed solar ratios 
within the current experimental uncertainties in the triple-${\alpha}$ 
reaction rate. Simultaneous variations by the same amount 
in both reaction rates or in either of them separately, however, can 
induce significant changes in the central $^{12}C$ abundance at core 
carbon ignition and in the mass of the supernova remnant.  
It therefore remains important to have experimental determinations 
of the helium burning rates so that their ratio and absolute values are 
known with an accuracy of 10\% or better.

\end{abstract}


\section{Introduction}

Throughout the past three decades much experimental and theoretical
effort has been dedicated to determining the rate of the
$^{12}C(\alpha,\gamma)^{16}O$ reaction.  It, and the triple-$\alpha$
($3\alpha$) reaction, are responsible for both energy generation and
the nucleosynthesis of C and O during stellar helium burning; the
ratio of their rates determines the ratio of carbon to oxygen at the
completion of core helium burning.  This ratio, in turn,
strongly influences the subsequent evolution of Type II supernova
(SNII) progenitors ($\gtrsim$ 9 \Msun stars), affecting both the
pre-supernova stellar structure and the post-explosive
nucleosynthesis.

Although progress has been achieved in the laboratory evaluation
of the $^{12}C(\alpha,\gamma)^{16}O$ reaction rate, $R_{\alpha,
12}$, there are significant uncertainties in its extrapolation to
the low energies relevant to hydrostatic helium-burning in stars
($\approx$ 300 keV). In a recent review, \cite{buc06} recommend
S(300 keV) = 145 keV b with errors in the range of 25\%-35\%
for the measured astrophysical S-factor of $^{12}C(\alpha,\gamma)^{16}O$. 
Stellar evolution calculations have shown (\citealt{wea93}; \citealt{boy02} 
[reported in \citealt{woo03}; \citealt{woo07}]) that such an 
uncertainty has major effects on SNII nucleosynthesis and on the mass 
of the pre-collapse core.

These calculations examined the changes in the production factors
(defined as the ratio of the average isotopic mass fraction of
nuclides in the ejecta to their solar mass fraction) induced by
varying $R_{\alpha,12}$. For these studies, however, the
\emph{pre-supernova} isotopic mass fractions were used in
determining the production factors. Under the crude assumption 
that SNII progenitors of close-to-solar metallicity are the main 
contributors to the observed solar abundances for the 
medium-weight isotopes (A = 16-40), very similar production 
factors are desirable for those nuclides. The reaction rate that 
produces the smallest spread in production factors was found; 
\cite{boy02} find the smallest spread for a narrow range 
in $^{12}C(\alpha,\gamma)^{16}O$ reaction rates that is only 10\% wide.

The calculations assumed a fixed  value of the $3\alpha$ reaction
rate, $R_{3\alpha}$. This is a reasonable assumption since $R_{3\alpha}$ 
has significantly smaller experimental  uncertainties, about 
10\%-12\% (\citealt{tur06}; \citealt{aus05}). However, if an accuracy 
of 10\% in the ratio $R_{3\alpha}$/$R_{\alpha,12}$
is required, the present accuracy of the $3\alpha$ rate is insufficient.

Besides uncertainties in nuclear reaction rates, uncertainties in the
initial isotopic composition of stars affect their evolution
and nucleosynthesis. Most recent studies of SNII evolution used
the abundances of \cite{and89};  to our knowledge there are no
systematic studies using the more recent abundance set of
\cite{lod03}.

In this paper we describe an extensive set of calculations to
determine how SNII nucleosynthesis and other stellar properties vary
when $R_{3\alpha}$ and $R_{\alpha,12}$ are varied. These calculations were
repeated for the two abundance sets: \cite{and89} and \cite{lod03}.
We repeated the calculations of \cite{boy02} to ensure that any 
small changes in procedures are unimportant. Another improvement 
is that the results of \cite{boy02}, as well as those of \cite{wea93},
were based on pre-SN nucleosynthesis, but some of the abundances
examined are known (\citealt{wea93}; \citealt{woo02}) to be modified
in the SN explosion.  The simulations presented in this paper include
explosive nucleosynthesis.

A description of the stellar models and the range of the
calculations is given in Section 2. The differences in the stellar
structure and nucleosynthesis resulting from differences in solar
abundance sets are presented in Section 3. In Section 4, we compare
the stellar evolution implications of the uncertainties in the $3\alpha$ and
$^{12}C(\alpha,\gamma)^{16}O$ rates.

\section{Computed models}

\begin{figure}
\centering
\includegraphics[angle=90,width=0.475\textwidth]{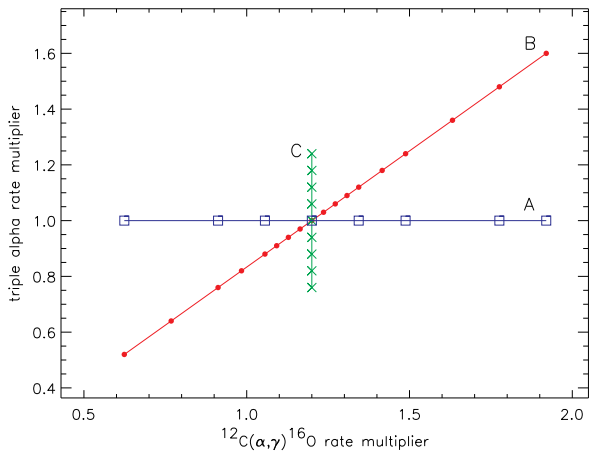}\\
\caption{Lines and dots show the three sets of simulations done
  for each star of a given initial mass and a given initial solar
  abundance distribution. For the blue squares, $R_{3\alpha}$ is held
  constant (at its value by \citealt{cau88}), and $R_{\alpha,12}$ is
  varied (A). For the red dots, both reaction rates are varied by the
  same percentage, so their ratio remains constant (B). For the green
  crosses, $R_{\alpha,12}$ is held constant at 1.2 times the rate
  recommended by \cite{buc96} and $R_{3\alpha}$ is varied (C).
}\label{range}
\end{figure}

Stars with initial masses from 13 to 27 \Msun were considered.  All
models were calculated using the implicit, one dimensional,
hydrodynamical stellar evolution code KEPLER. Since its first
implementation in 1978 (\citealt{wea78}) the code
has undergone several major revisions with improvements to the
physical modeling of the stellar structure and to the nuclear 
reaction networks (\citealt{woo95}; \citealt{rau02}; \citealt{woo02}).  
A small network directly coupled to the stellar model 
calculation provides the approximate nuclear
energy generation rate. A larger ``adaptive'' network is used to 
track nucleosynthesis. The large network automatically adjusts 
its size to accommodate the current nuclear flows and progressively 
grows from several hundred isotopes during hydrogen burning 
to more than 2000 isotopes at explosive burning. The treatment of convection,
semi-convection, and overshoot mixing is as described in \cite{woo88} 
and \cite{woo02}. We do not include the effects of rotation and 
magnetic fields. Stars are first evolved from
the zero age main sequence to pre-supernova, i.e., from central
hydrogen burning to iron core collapse, and are then exploded.  The
explosion is parameterized by a piston at a constant Lagrangian mass
coordinate and has two important specifications: its location in mass
(i.e., the initial mass cut) and the total kinetic energy of the
ejecta at infinity (here, 1 yr after the explosion). See \citet{woo07} 
for a more complete description of the piston parameters.

The values of these two parameters are chosen to fit reasonably well
within the range of observational constraints. The explosion energy
was set to 1.2B (B for Bethe, $1\,\mathrm{B}=10^{51}$ ergs).
Supernova 1987A is thought to be an 18-20\,\Msun star which exploded
with an estimated energy in the range 0.6-1.5B with an uncertainty of
perhaps 50$\%$ based on the observed light curve and velocity
(\citealt{arn89}).  The initial mass cut was placed at the base of the
oxygen burning shell, a location associated with a large density drop,
and hence dynamically important to generating successful explosions
(\citealt{jan07}). Specifically, we chose to place the piston at the 
location in the star where the entropy S reaches a value of $S=4 k_B$/baryon 
(\citealt{woo07}), beyond which a large rise in entropy, hence a drop in 
density, is observed. The piston location cannot be 
below the surface of the iron core or neutron-rich species in the 
iron group will be overproduced; it cannot be above the base of the 
oxygen shell or typical neutron star masses will be too 
large (\citealt{woo07}). Our nucleosynthesis studies do take into 
account all strong and weak reactions during oxygen shell burning, 
including the slight neutron excess resulting in this burning phase.
KEPLER calculations by \cite{woo07} showed that explosion energies 
of either 1.2B or 2.4B and mass cuts at the base of the oxygen burning 
shell or at the edge of the iron core gave very similar nucleosynthesis, 
except for the iron peak nuclei. We note, however, that in a recent 
study by \cite{you06}, both elemental and isotopic yields beyond 
silicon were found to be very sensitive to the explosion energy.

\begin{figure}
\centering
\includegraphics[angle=90,width=0.475\textwidth]{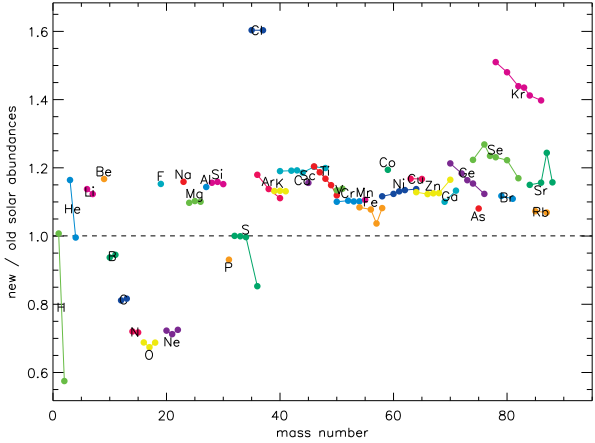}\\
\caption{Ratio of the \cite{lod03} abundances to the \cite{and89}
abundances as a function of mass number (up to strontium).
Isotopes of each element have the same color and are connected by
lines.}\label{aburatio}
\end{figure}

\begin{figure*}
\centering
\includegraphics[angle=90,width=0.475\textwidth]{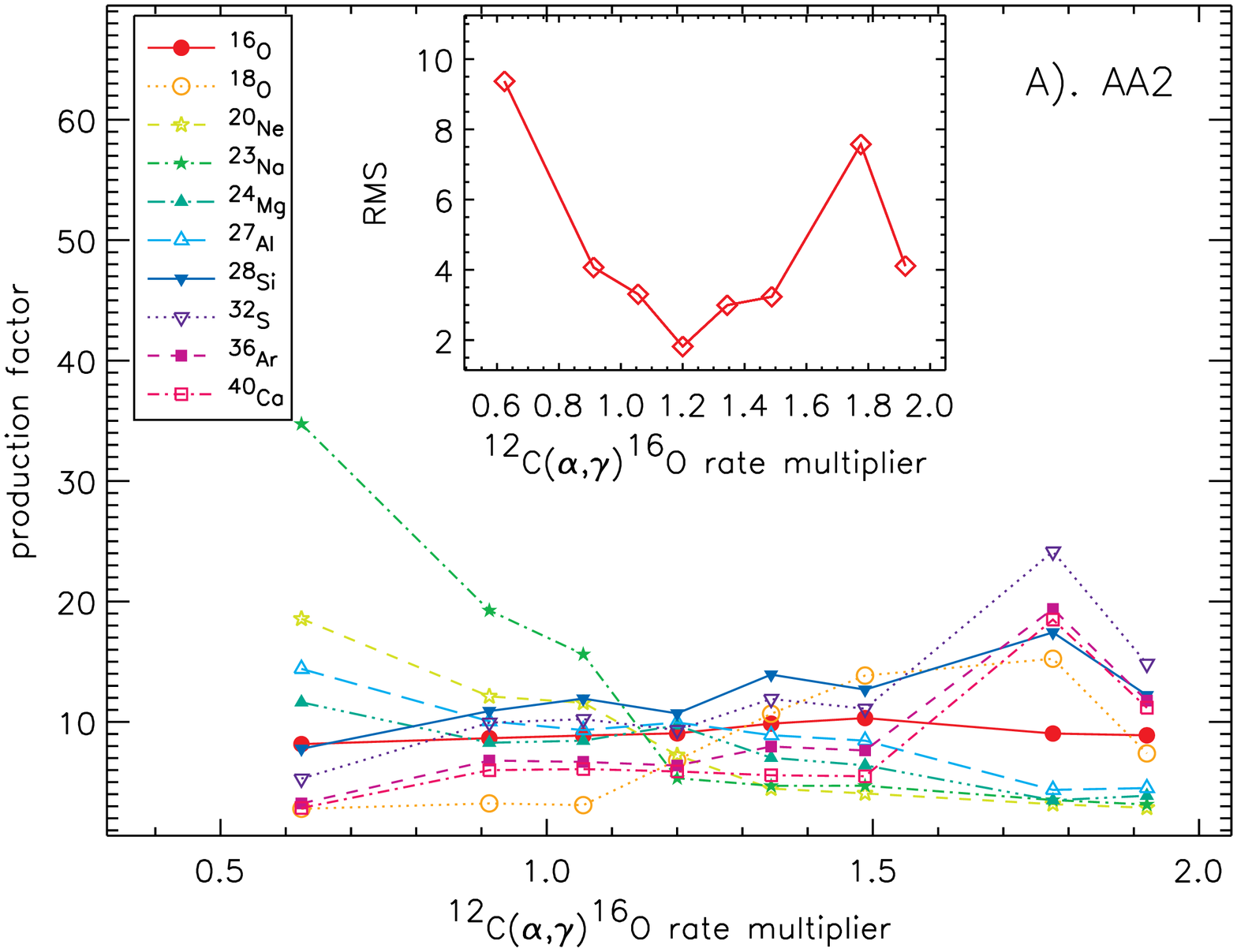}
\hfill
\includegraphics[angle=90,width=0.475\textwidth]{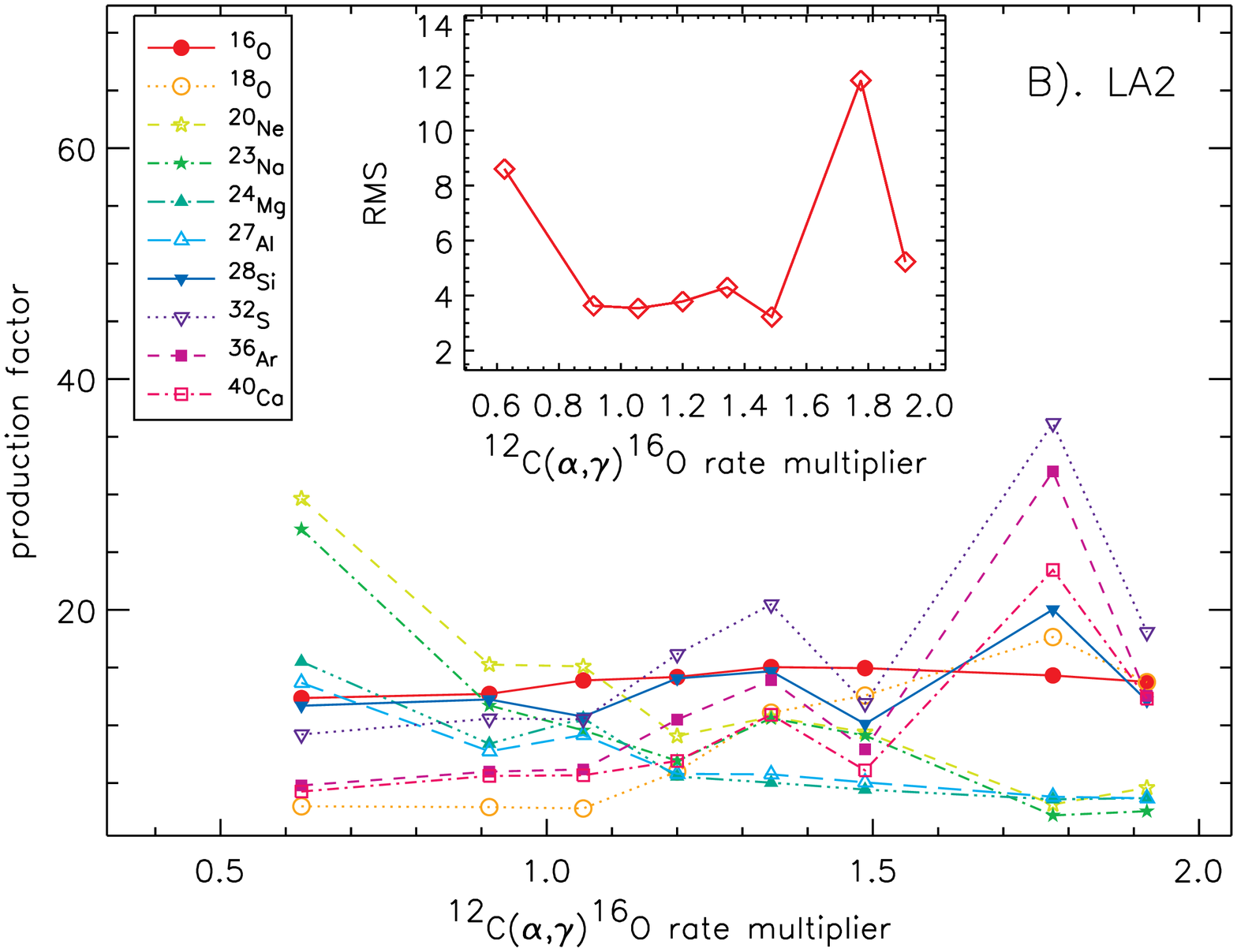}
\\
\includegraphics[angle=90,width=0.475\textwidth]{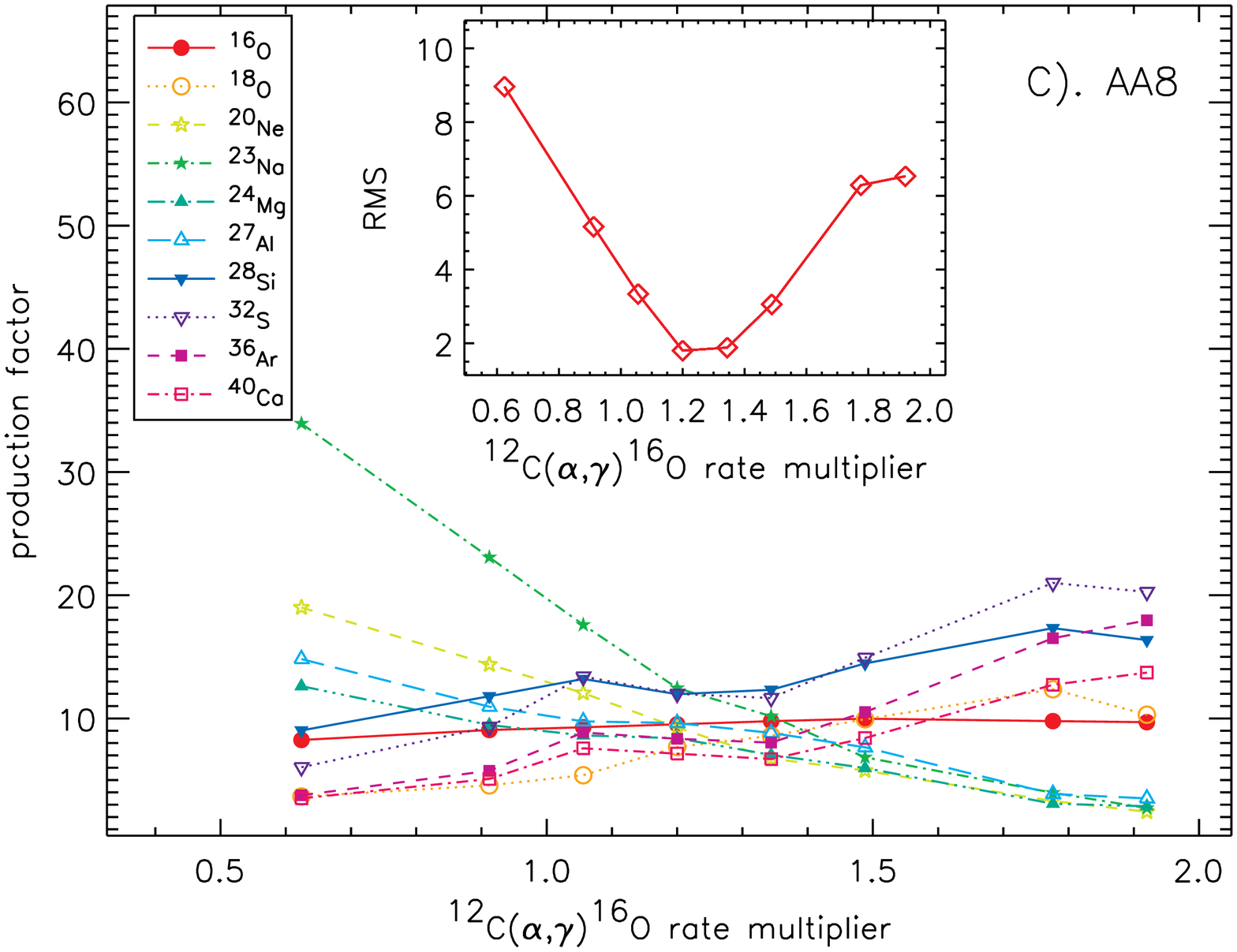}
\hfill
\includegraphics[angle=90,width=0.475\textwidth]{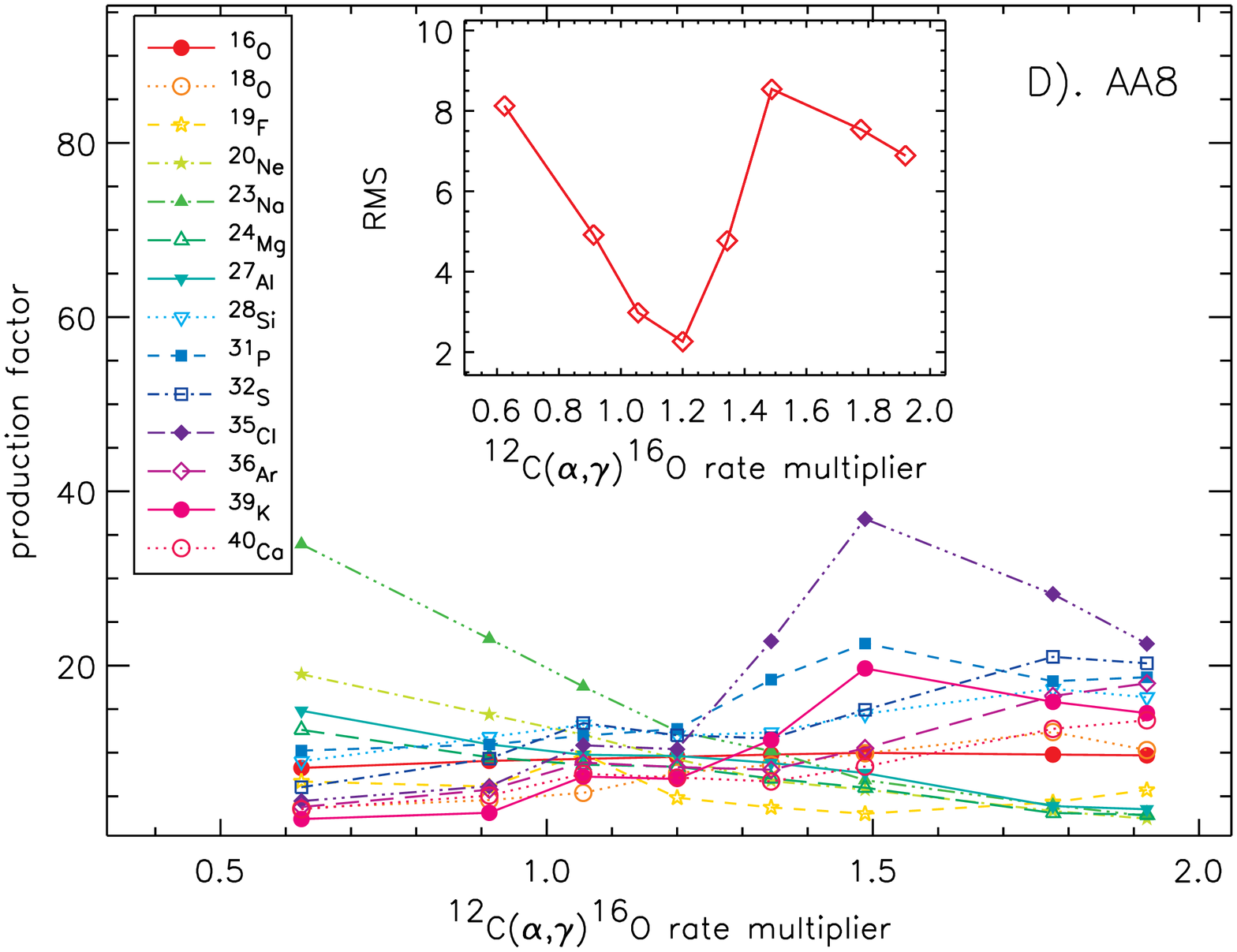}
\caption{\textbf{Panel A:} Production factors for AA2 and their rms deviations 
from the mean for the same set of isotopes as those selected by \cite{boy02}. 
A multiplier of 1 means a rate of 1 times the rate recommended 
by \cite{buc96}. 
\textbf{Panel B:} Same as Panel A, but for LA2.
\textbf{Panel C:} Same as Panel B, but for AA8
\textbf{Panel D:}
Same as Panel C, but with the addition of the production factors for 
$^{19}$F, $^{31}$P, $^{35}$Cl, and $^{39}$K.} \label{pfconst3a}
\end{figure*}

\begin{figure*}
\centering
\includegraphics[angle=90,width=0.475\textwidth]{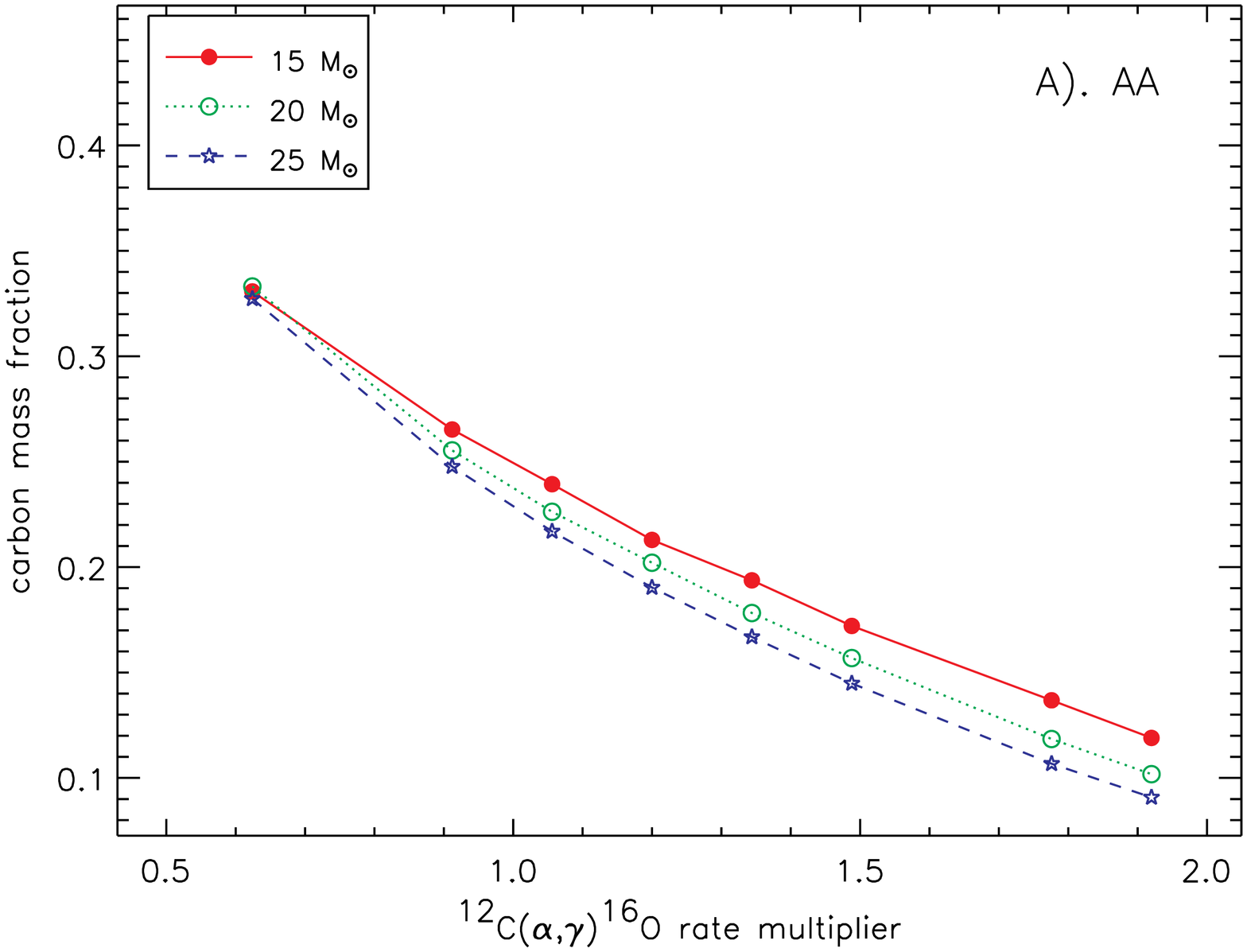}
\hfill
\includegraphics[angle=90,width=0.475\textwidth]{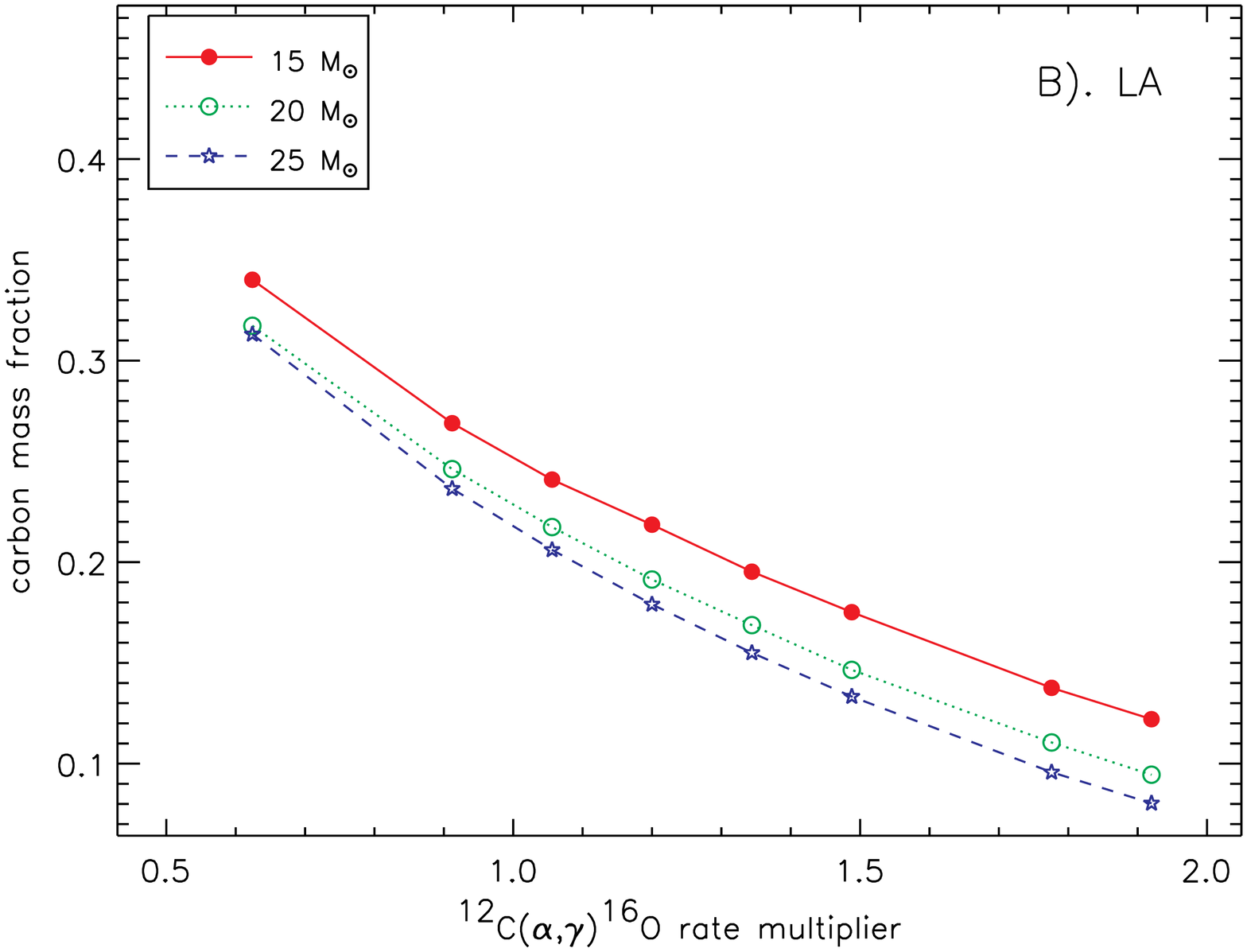}
\\
\includegraphics[angle=90,width=0.475\textwidth]{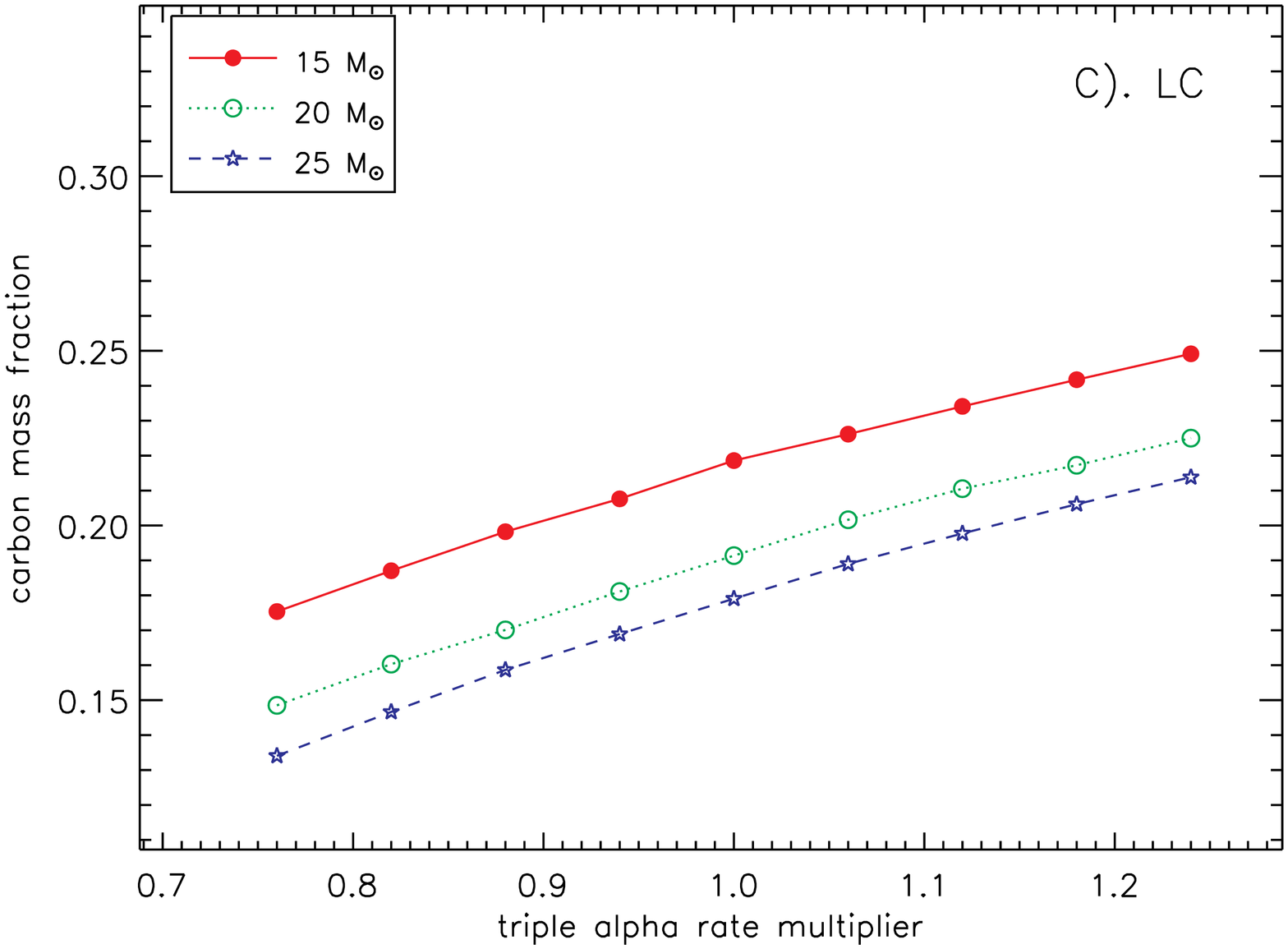}
\hfill
\includegraphics[angle=90,width=0.475\textwidth]{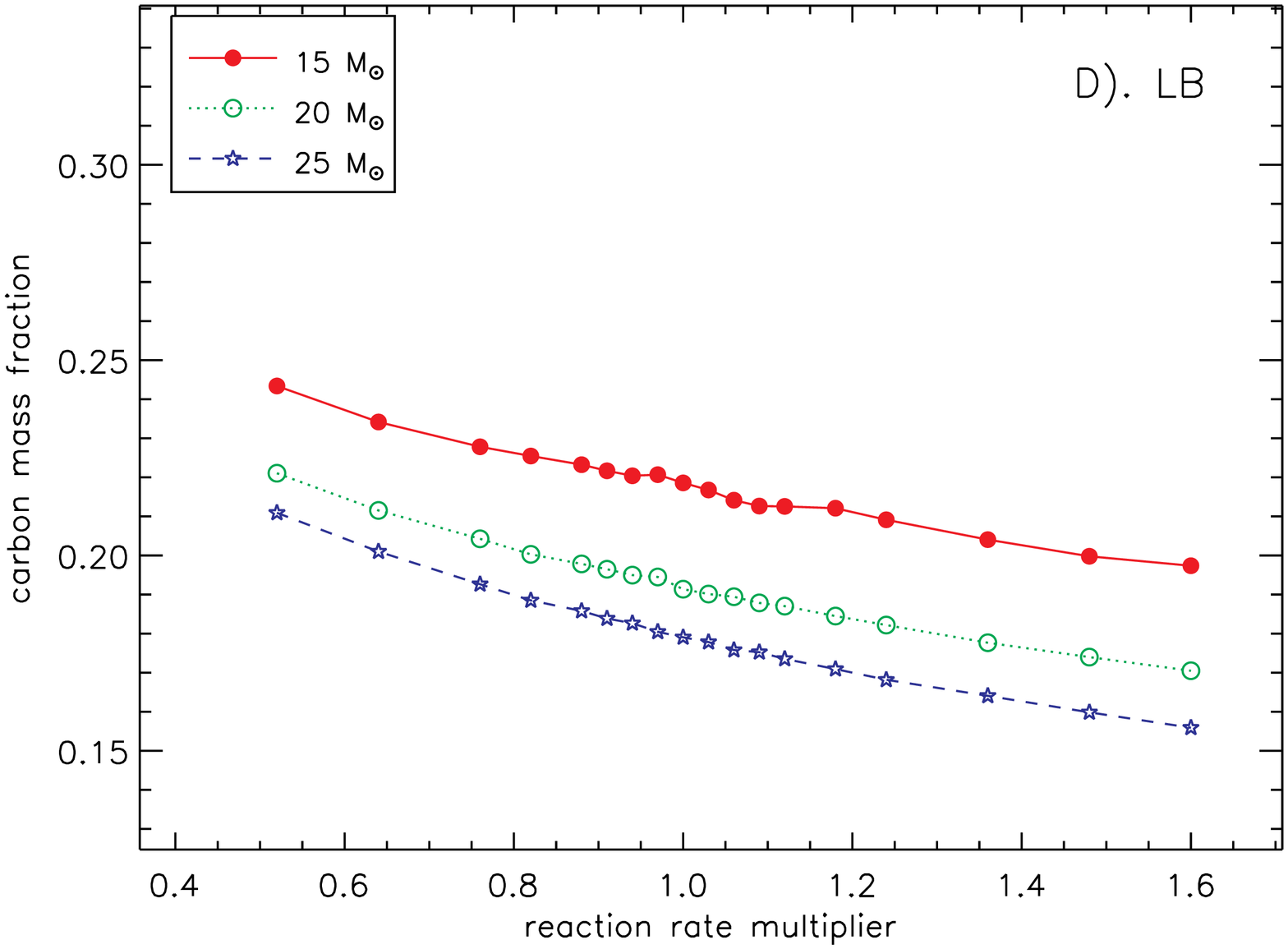}
\caption{\textbf{Panel A:} Carbon mass fraction at the center of
the star at core carbon ignition for 15, 20, and 25 \Msun stars and 
for the AA series.
\textbf{Panel B:} Same as Panel A, but for LA.
\textbf{Panel C:} Same as Panel A, but for LC.
\textbf{Panel D:} Same as Panel A, but for LB.} \label{C12abu}
\end{figure*}

\begin{figure*}
\centering
\includegraphics[angle=90,width=0.475\textwidth]{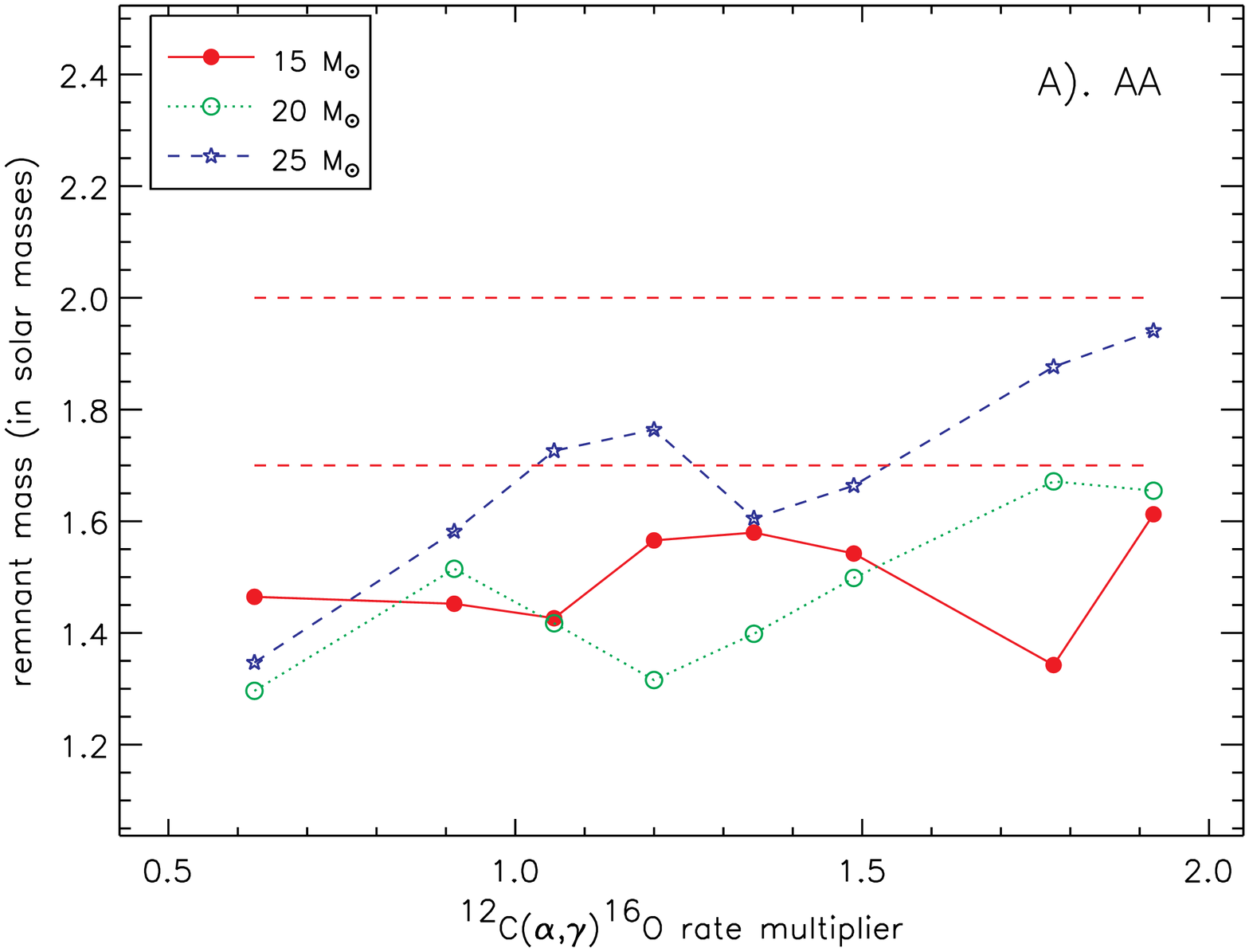}
\hfill
\includegraphics[angle=90,width=0.475\textwidth]{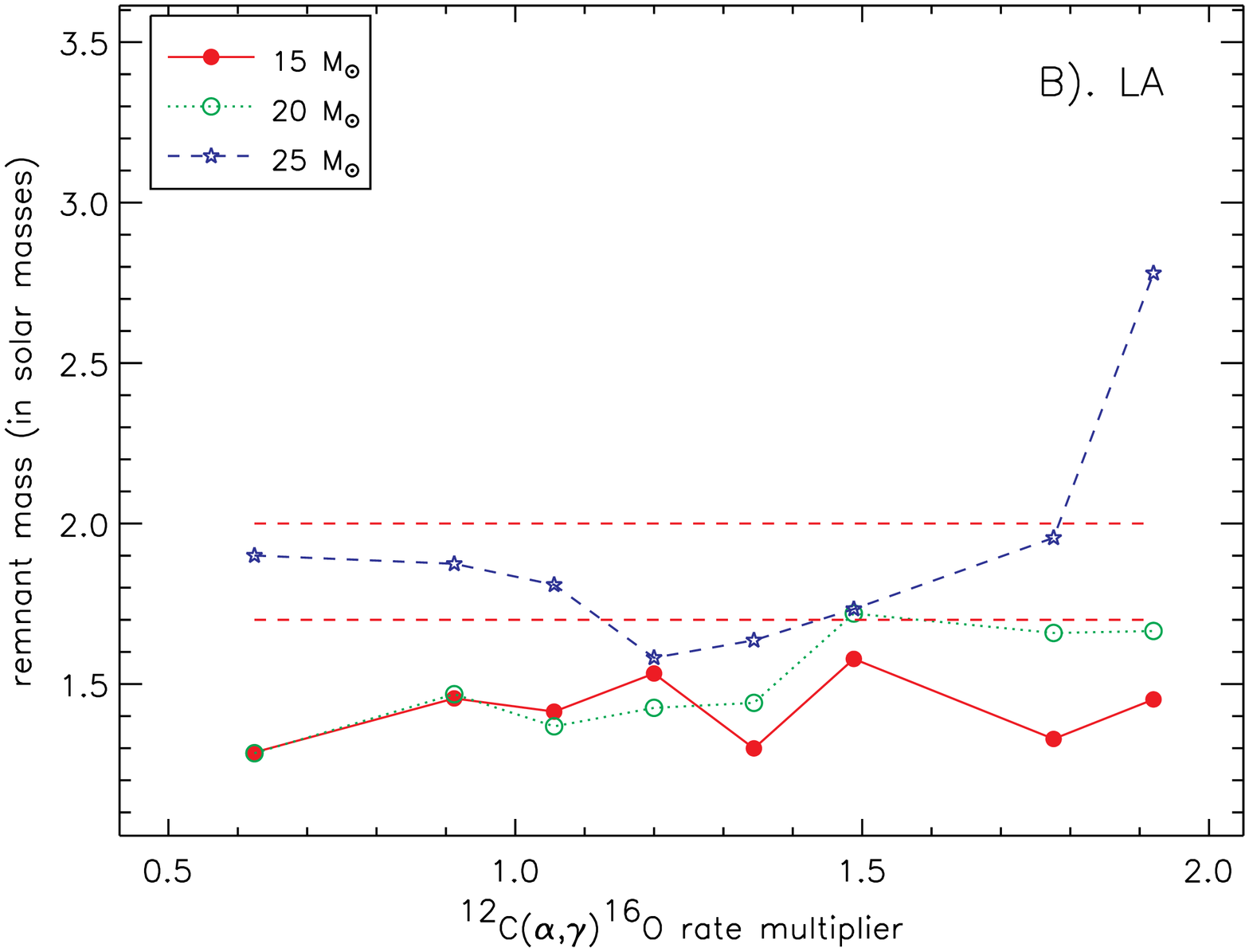}
\\
\includegraphics[angle=90,width=0.475\textwidth]{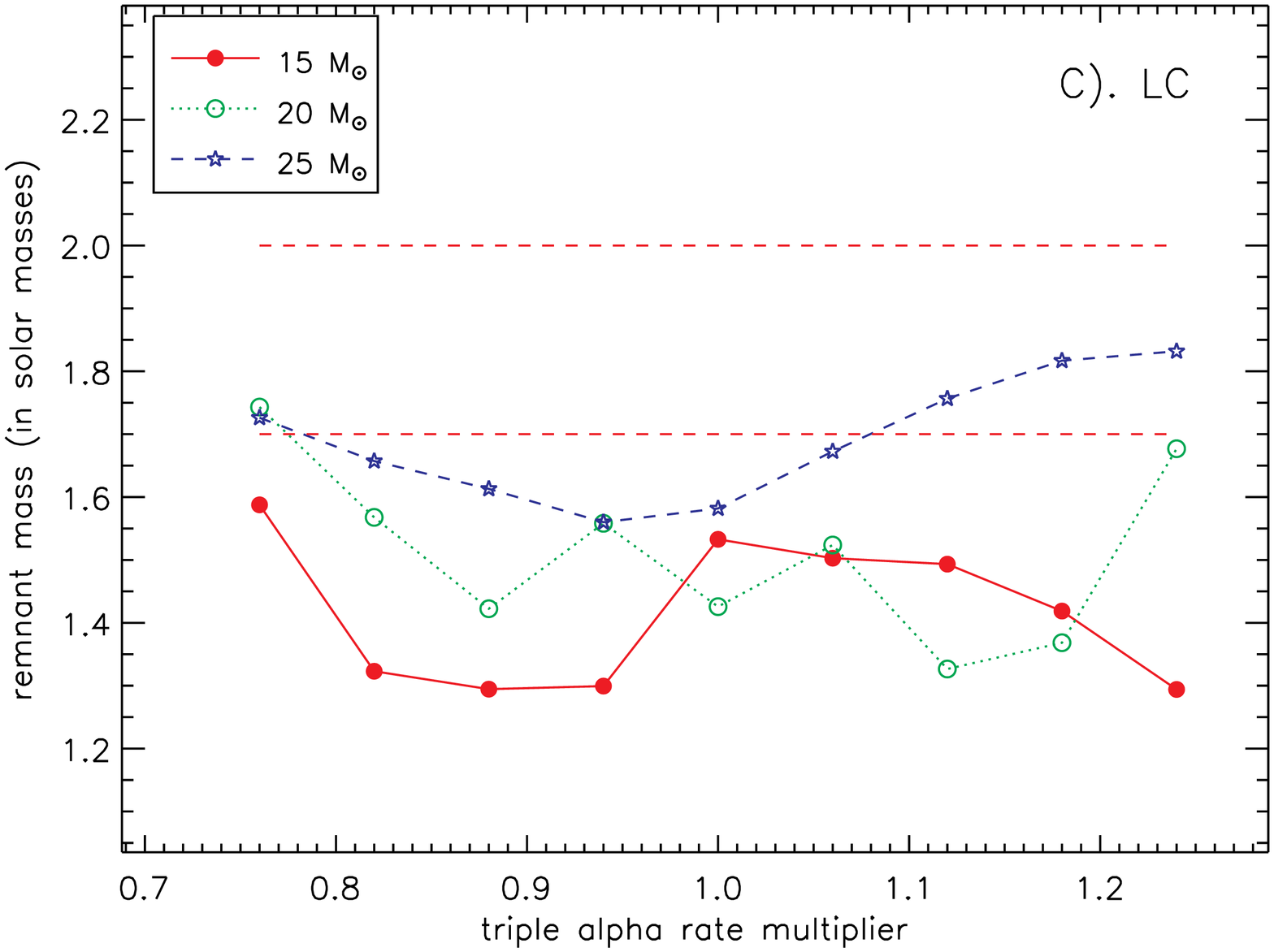}
\hfill
\includegraphics[angle=90,width=0.475\textwidth]{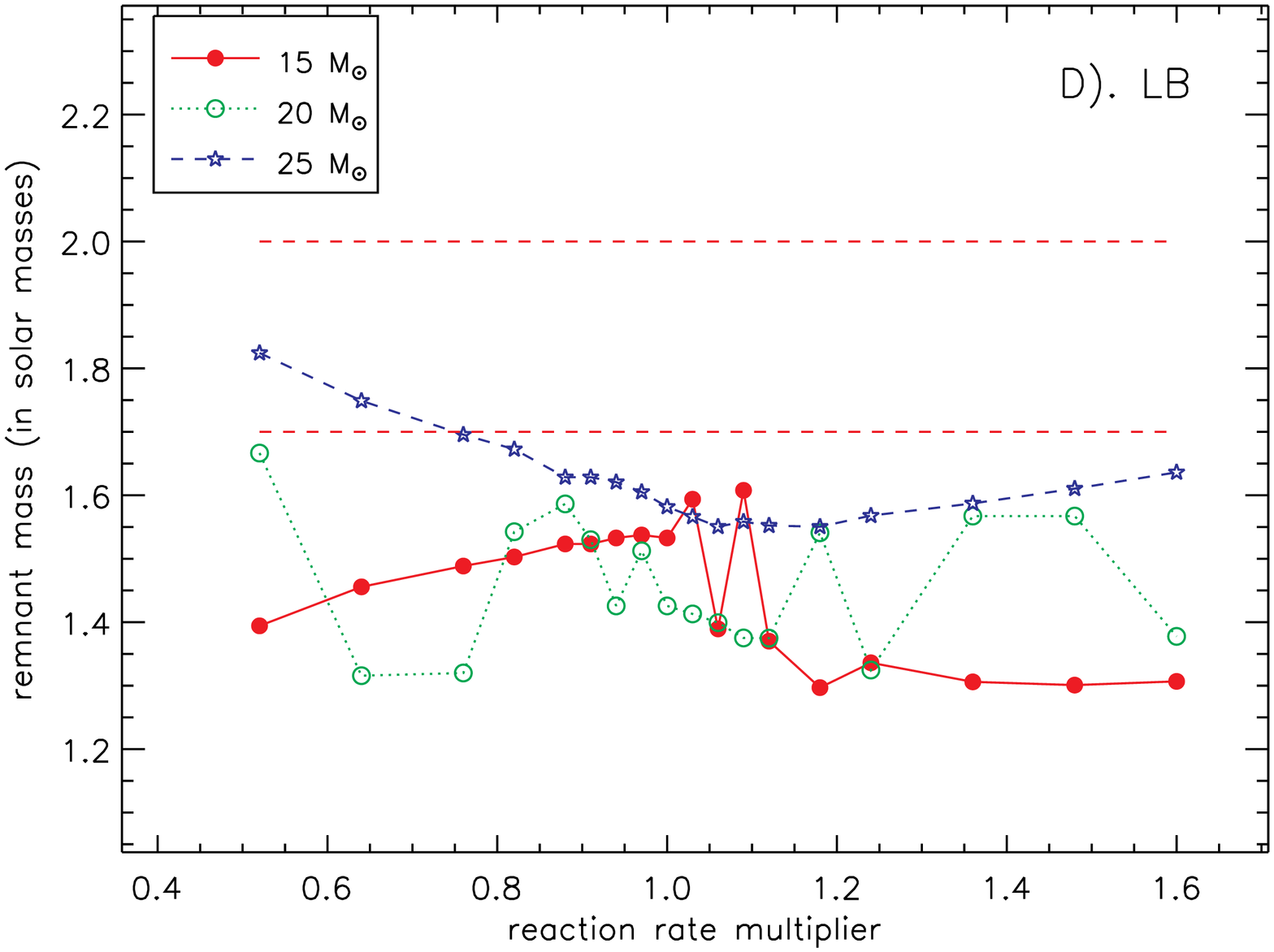}
\caption{Gravitational mass of the remnant (neutron star or black hole) 
after explosion for 15, 20, and 25 \Msun stars. The dotted lines 
at 1.7 and 2.0 \Msun mark possible maximum masses for neutron 
stars (\citealt{lat07}); for heavier masses, black holes may be formed.
\textbf{Panel A:} AA series.
\textbf{Panel B:} LA series.
\textbf{Panel C:} LC series.
\textbf{Panel D:} LB series.} \label{remnants}
\end{figure*}

Three separate studies were done for stars of 15, 20, and 25 \Msun for
both solar abundance sets, \cite{and89} and \cite{lod03}: (A)
$R_{3\alpha}$ was kept constant (at its value from \citealt{cau88})
and $R_{\alpha,12}$ was varied; (B) both rates were varied by the same
factor, so their ratio remained constant; and (C) $R_{\alpha,12}$ was
held constant at 1.2 times the rate recommended by \cite{buc96} and
$R_{3\alpha}$ was varied. The ranges of those variations are shown in
Figure~\ref{range}. For the \cite{and89} abundances, we additionally
computed stars of 13, 17, 19, 21, 23 and 27 \Msun to have a better
sampling of the initial mass function (IMF) (see eq. \ref{IMF}),
in order to better integrate over intrinsic star to star variations
and thereby reduce the impact of numerical noise in the production factors.
The isotopic mass fractions from all the stars in a given study were
then averaged over an IMF with a slope of $\gamma=-2.6$ (\citealt{sca86}) and
divided by their solar mass fraction, giving the production factor of
each isotope. The slope $\gamma$ is defined by the equation:
\begin{equation}
\xi(log M) \approx AM^{\gamma}\label{IMF}
\end{equation}
where $\xi(log M)$ is the IMF defined in units of
the number of stars per (base 10) logarithmic mass interval $M$ per 
square parsec of the Galactic disk, $M$ is the initial mass of the 
star in solar masses, and $A$ and $\gamma$ are constants (\citealt{wea93}).

\begin{deluxetable}{ll}
\tablecaption{Simulation series \label{series}}
\tablehead{\colhead{Label} & \colhead{Description}}
\startdata
AAX\tablenotemark{a} & \cite{and89}; $R_{\alpha,12}$ varied \\
ABX\tablenotemark{a} & \cite{and89}; $R_{3\alpha}$, $R_{\alpha,12}$ varied \\
ACX\tablenotemark{a} & \cite{and89}; $R_{3\alpha}$ varied \\
\\
LAX\tablenotemark{a} & \cite{lod03}; $R_{\alpha,12}$ varied \\
LBX\tablenotemark{a} & \cite{lod03}; $R_{3\alpha}$, $R_{\alpha,12}$ varied \\
LCX\tablenotemark{a} & \cite{lod03}; $R_{3\alpha}$ varied \\
\enddata
\tablenotetext{a}{X=2, if IMF average over 2 stars (15 and 25 \Msun); X=8, if IMF average over 8 stars (13, 15, 17, 19, 21, 23, 25, and 27 \Msun)}

\end{deluxetable}

We will adopt a three-character notation to label our plots, e.g., 
LA2 (see Table \ref{series}). The first character can be an L (to denote the
\citealt{lod03} initial abundances) or an A (for the \citealt{and89}
initial abundances).  The second character denotes the study: A when 
$R_{3\alpha}$ was kept constant and $R_{\alpha,12}$ was varied; B 
when both rates were varied by the same factor, so their ratio 
remained constant; and C when $R_{\alpha,12}$ was held constant 
and $R_{3\alpha}$ was varied.  The third character is a number; it is 2 
when the production factors are averaged over two stars ($15\,\Msun$ and
  $25\,\Msun$) and 8 when the average is over eight stars ($13$, $15$, $17$, 
$19$, $21$, $23$, $25$, and $27\,\Msun$). When no third character is 
present, no average has been performed, as in the case where the 
numbers only apply to a single star.

\section{Sensitivity to differences in solar abundances and reaction rates}

The differences in the two recent solar abundance
determination are shown in Figure \ref{aburatio}. For \cite{lod03}
compared to \cite{and89}, the abundances are the following: for CNO they 
 are lower by about 30$\%$; for Cl, Kr, Xe, and Hg they are higher by 
more than 40$\%$; and for most other metals they are higher by about 15$\%$. 
As a consequence, the overall solar mass fractions change from 
$X_{0} = 0.7057$, $Y_{0} = 0.2752$ and $Z_{0} = 0.0191$ for the old set to 
$X_{0} =0.7110$, $Y_{0} = 0.2741$, and $Z_{0} = 0.0149$ for the new set.

\subsection{The effect on the production factors}

The study with $R_{3\alpha}$ constant and $R_{\alpha,12}$ varied, is
an elaboration of two previous studies using the \cite{and89}
abundances (\citealt{wea93}; \citealt{boy02}). Relative to
\cite{wea93}, our models also include mass loss due to stellar winds,
as described in \cite{woo07}. As noted above,
explosive nucleosynthesis is also included. The same study done with the
\cite{lod03} solar abundances is entirely new and demonstrates the
uncertainties in determining $R_{3\alpha}$ and $R_{\alpha,12}$ using
astrophysical models.

Based on SNII nucleosynthesis considerations, \cite{wea93} predicted
an S-factor at 300 keV of $\sim$170 keV b, or more precisely a rate
of 1.7 $\pm$ 0.5 times that of \cite{cau88}. This constrained 
$R_{\alpha,12}$ to a range of about 30\%. The same study was 
repeated later by \cite{boy02} (reported in \citealt{woo03}; 
\citealt{woo07}) with improved stellar models [newer opacities, 
added mass loss, finer stellar zoning, and finer grid
of $^{12}C(\alpha,\gamma)^{16}O$ rates] and found a best fit of 175
keV b or about 1.2 times the value of S(300 keV) suggested by
\cite{buc96} (146 keV b). This study concluded that $R_{\alpha,12}$ needed
to be known to $\leq 10\%$ (\citealt{woo03}; \citealt{woo07}).

In Figures \ref{pfconst3a}A, \ref{pfconst3a}C and \ref{pfconst3a}D,
we illustrate our results for the \cite{and89} abundances. Figure
\ref{pfconst3a}A shows the production factors averaged over two
stars (AA2), and their rms deviations for the same set
of isotopes as those selected by \cite{boy02}. Figure \ref{pfconst3a}C
does the same for a larger group of stars (AA8) and 
Figure \ref{pfconst3a}D does the same for a larger
set of medium-weight isotopes (now including $^{19}$F, $^{31}$P,
$^{35}$Cl, and $^{39}$K). If SNII are indeed the major site of
production of all medium-weight elements (A = 16-40), then those
elements should have similar production factors at a point where
their rms deviations are minimum. For the \cite{and89} abundances,
the conclusion seems robust; the position of the minimum is well
defined at a rate of 1.2 times the \cite{buc96} rate for different
sets of stars and nuclides, although the details of the rms curves
vary somewhat. This conclusion agrees with the earlier work by
\cite{wea93} and \cite{boy02}.

For the \cite{lod03} initial abundances the results are less
definitive, as shown in Figure \ref{pfconst3a}B. The average
production factors at the minimum are about the same for both
abundance sets. However, the rms curve now has a much \emph{broader}
minimum, again centered around 1.2 times the \cite{buc96} rate, but
extending from a rate multiplier of 0.9 to 1.5. The spread in
production factors at the minimum is larger by about a factor of
2. These production factors apparently provide a much less
stringent constraint on $R_{\alpha,12}$, allowing a range of
$\pm$25$\%$ around the central value of 1.2 times the \cite{buc96}
rate. This is unfortunate, since it means that one cannot so
strongly limit the uncertainty in $R_{\alpha,12}$ using SNII
calculations of production factors.

\begin{figure*}
\centering
\includegraphics[angle=90,width=0.475\textwidth]{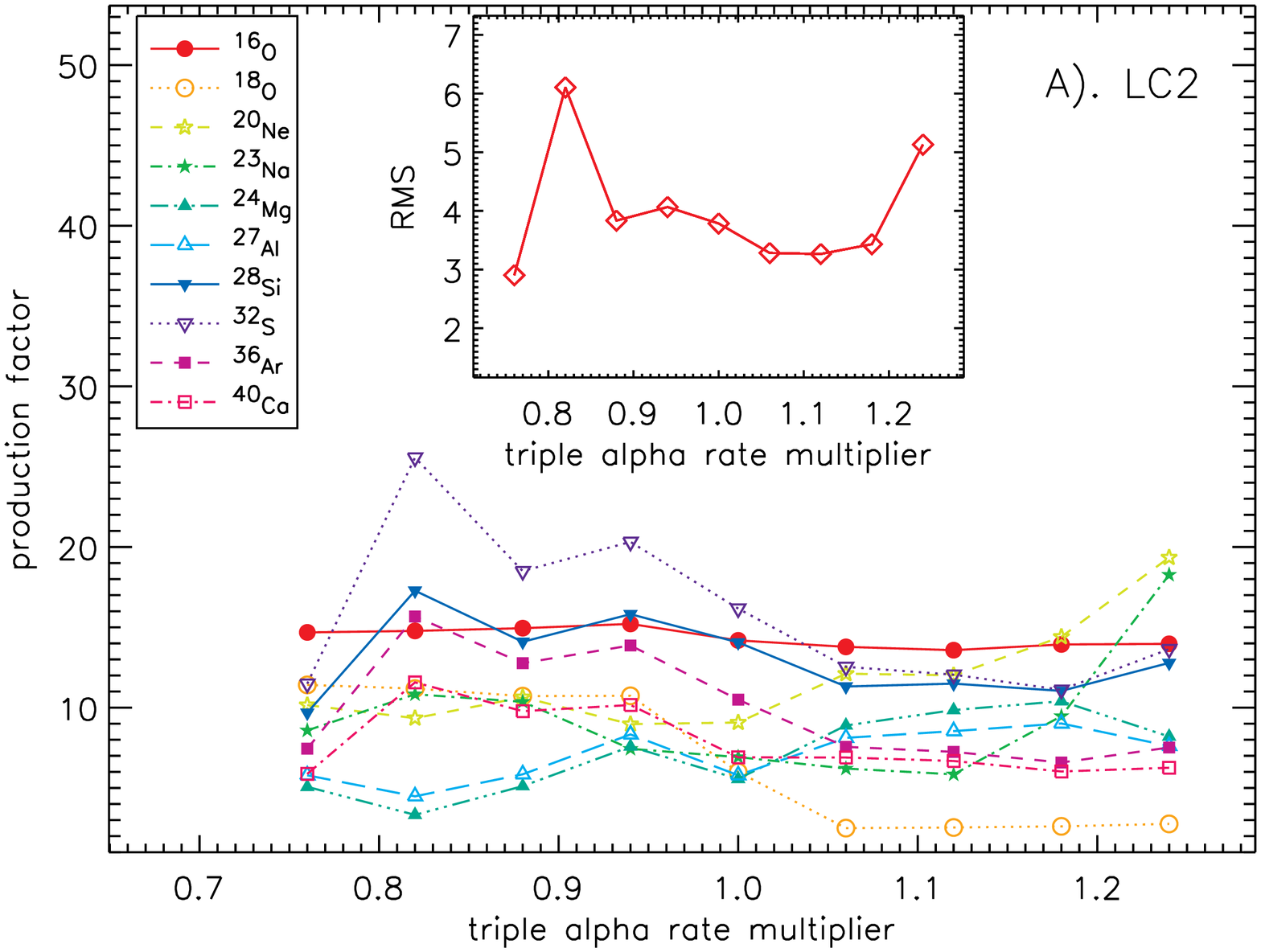}
\hfill
\includegraphics[angle=90,width=0.475\textwidth]{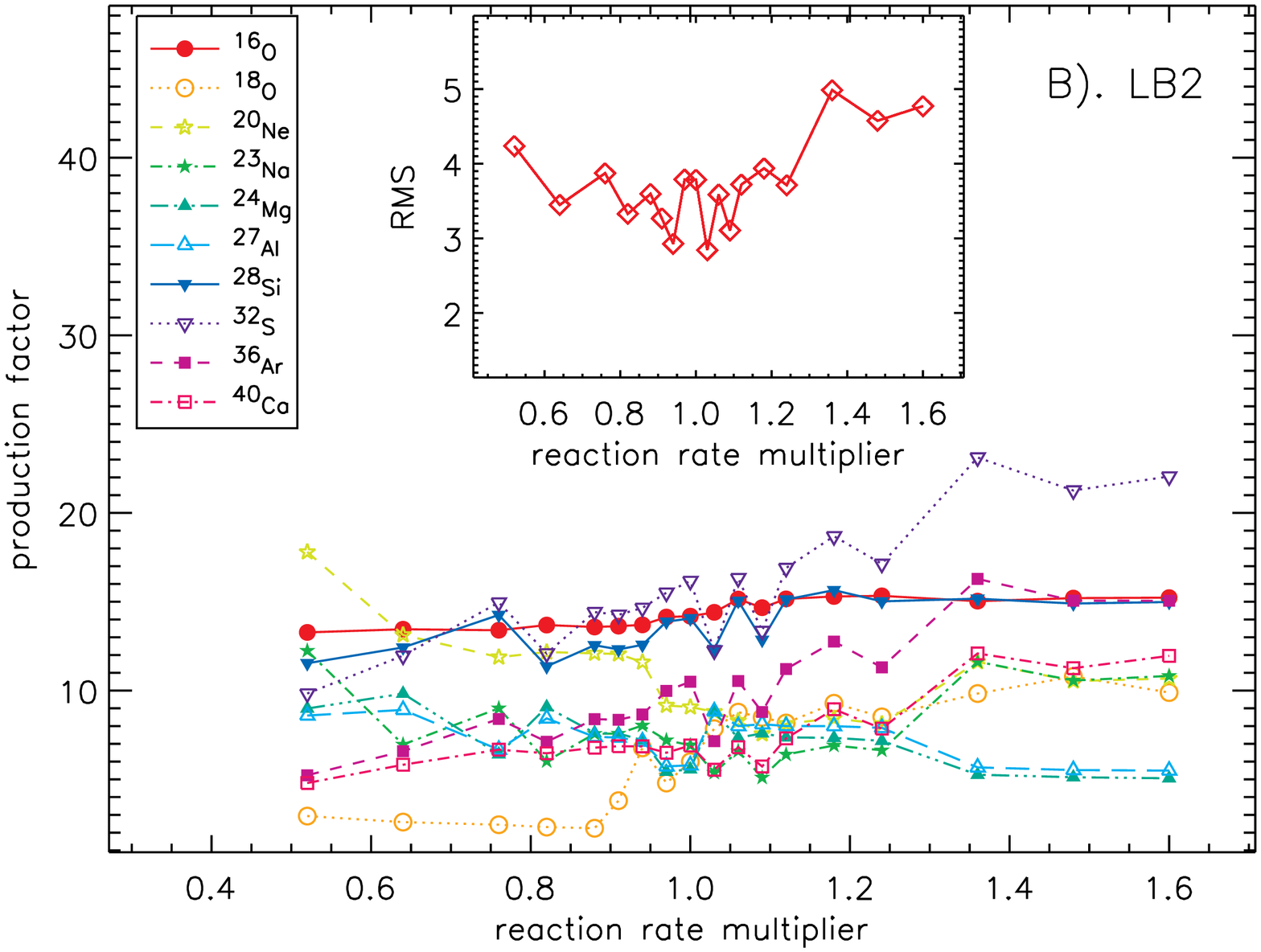}
\caption{\textbf{Panel A:} Production factors and their rms deviations 
from the mean for some medium-weight isotopes (the same set 
as \citealt{boy02}) for LC2.
\textbf{Panel B:} Same as Panel A, but for LB2.}
\label{Lodderspf}
\end{figure*}

Our results are for post-explosion values of the production factors,
whereas the two previous studies stopped at the pre-supernova stage.
We found that nuclides beyond $^{28}$Si, such as $^{40}$Ca,
$^{36}$Ar and $^{32}$S, were significantly modified during
explosion, often by a factor of 1.5 or more for the \cite{and89}
abundances.  Yet,  these modifications did not greatly change the
earlier results for production factors.

The 20 \Msun star showed a peculiar behavior: large
over-productions were found for $^{31}$P, $^{35}$Cl, and $^{39}$K
for some choices of $R_{3\alpha}$ and $R_{\alpha,12}$. This was also 
observed and explained in \cite{rau02}.  The
 over-productions are attributed to the merging of the
convective oxygen-, neon- and carbon-burning shells about 1 day
before the explosion, thereby carrying neutron sources such as
$^{22}$Ne and $^{26}$Mg to depths where they burn rapidly and
provide neutrons for capture reactions. Because of these
peculiarities, we excluded the 20 \Msun star from our results.

\subsection{Variations in the carbon mass fraction at central carbon ignition and in the remnant masses}

We also explored the change of the central carbon mass fraction at
core carbon ignition and of the remnant mass after explosion, both as
a function of the initial solar abundances of the stars and as a
function of variations in the helium-burning reaction rates. We
illustrate the results in Figures 4 (carbon mass fraction), and 5
(remnant masses). The remnant masses are the gravitational masses of
the resulting neutron stars or black holes. They are based on the
baryonic mass below the piston (i.e., the mass enclosed within a
radius reaching out to the base of the oxygen shell at the
pre-supernova stage) corrected for the binding energy
(\citealt{zha07}) according to the approximation given by
\cite{lat01}. In our study, none of the stars had any significant 
fallback after explosion.

The variations in the central carbon mass fractions
are smooth, but we see a very sensitive dependence of the remnant
masses on the solar abundance set used for the initial stellar
composition. To disentangle and assess the magnitude of the effects 
compared to observational data, however, would require a detailed 
population synthesis study of remnant masses as a function of 
metallicity which is beyond the scope of this paper. The predicted 
remnant masses are also strongly dependent on the precise reaction 
rates used in the pre-supernova evolution, often  
varying by 0.2 \Msun or more, over ranges of $\pm 2\sigma$ experimental 
errors of the reaction rates.

The remnant mass may determine the relative population of
neutron stars and black holes resulting from SNII explosions. \cite{lat07}
have surveyed the available data on neutron star masses.  Their conclusions 
are the following:  (1)  While some masses in excess of 2 \Msun have been 
reported, ``it is furthermore the case that the 2$\sigma$ errors for 
all but two systems extend into the range below 1.45 \Msun, so caution 
should be exercised before concluding that firm evidence of large 
neutron star masses exists.'' And (2) the smallest ``reliably estimated 
neutron star mass is 1.18 $\pm 0.02$  \Msun''.  While these uncertainties 
make it difficult or impossible to use neutron star masses to place limits 
on reaction rates or abundances, for  orientation we have placed lines 
in  Fig. 5, at values of 1.7 and 2.0 \Msun as possible maximum masses 
for neutron stars.

\section{Comparing changes in triple-$\alpha$ and $^{12}C(\alpha,\gamma)^{16}O$ rates}

In this section, we discuss the relative importance of the
uncertainties in  $R_{3\alpha}$ compared to the 2 times
larger uncertainties in $R_{\alpha,12}$.
Figures \ref{Lodderspf}A and \ref{Lodderspf}B show the
production factors of some medium-weight isotopes (the same set
as \citealt{boy02}) as a function of the triple-$\alpha$ reaction rate
in two of our studies: $R_{3\alpha}$ varied and
$R_{\alpha,12}$ constant, and both reaction rates varied by the 
same factor. The variations in the production factors 
(Figure \ref{Lodderspf}A) over a range of one
standard deviation $\sigma$ ($3\alpha$ multiplier from 0.88 to
1.12) are small, although there are larger deviations for
2$\sigma$ differences.

We find a very sensitive dependence of the remnant masses on the
helium burning reaction rates, and on the initial solar abundance set
used. The smooth decrease in the carbon mass fraction as a function of
increasing $R_{\alpha,12}$, or decreasing $R_{3\alpha}$, is expected.
The following argument is commonly given to explain the general
increasing trend of the remnant masses when the $R_{\alpha,12}$ is
increased (seen in Figure \ref{remnants}A for instance): a smaller
rate gives a larger carbon abundance after helium burning. During
carbon shell burning, this larger abundance supports longer and more
energetic burning which allows the central regions to cool to lower
entropy. The lower entropy, in general, gives smaller iron cores
(hence remnants) for stars of a given main-sequence mass
(\citealt{woo03}). Figure \ref{C12abu} also shows that smaller 
stars make more carbon than larger ones, reflecting 
their higher density, and tend to have smaller remnants
following explosion (as seen in Figure \ref{remnants}), which supports  
the previous argument. When looking at the remnant masses
for the $25\,\Msun$ star (Figure \ref{remnants}C) the same argument
seems to break down at least partly.  One expects a general decreasing
trend in the remnant masses for higher triple-$\alpha$ rates, whereas one
sees an increase for a multiplier larger than one. The non-monotonic
behavior of remnant masses can be understood as
a result of the interaction of subsequent burning shells.  This causes the
behavior of the remnants for the 25 \Msun star of Figure
\ref{remnants}C.

In Figure \ref{remnants}, variations within the current experimental range of
uncertainties ($2\sigma$) of both $R_{3\alpha}$ and $R_{\alpha,12}$
cause significant changes in the remnant mass.  The remnant mass
curves look smooth for the 25 \Msun star, but, they show an
oscillatory behavior with rapid variations (over a small rate
multiplier range) for our 15 and 20 \Msun stars. In particular, Figure
\ref{remnants}D shows very strong fluctuations in remnant masses, when
the ratio of the helium burning reactions is kept constant (LB),
despite the very smooth change of the carbon mass fractions. These
oscillations are likely due to small numerical noise in the models
originating from temporal and spatial discretization, combined with a
sharp transition in the stellar evolution past helium burning as a
function of the carbon mass fraction, where an additional burning
shell ignites or does not ignite beyond a certain threshold.

These observations lend support to the idea that variations 
in both $R_{3\alpha}$ \emph{and} $R_{\alpha,12}$ are important, not
just their ratio or their relative variations. An increase of 10\%
in $R_{3\alpha}$ gives the same amount of increase in the
central carbon mass fraction as an 8\% decrease in $R_{\alpha,12}$, in 
close agreement with the findings of \cite{woo07} for a simple 
calculation at given temperatures and densities. A 27\% decrease 
in both reaction rates is required to produce the same amount 
of increase in the central carbon mass fraction when the two 
rates are multiplied by the same factor.

\section{Conclusion}

Our simulations show that multiple uncertainties significantly
influence the evolution and nucleosynthesis of SNII in current one dimensional
massive star and supernova models. The notable effect of differences
in solar abundance sets is one example. Using the \cite{lod03}
abundances rather than the previous standard set by \cite{and89},
appears to reduce the precision with which SNII simulations of
production factors can be used to constrain
$R_{\alpha,12}$ to $\pm25\%$. The production factors of
medium-weight elements (A = 16-40) were found to be about constant
within the current $1\sigma$ experimental uncertainties in the triple-$\alpha$
reaction rate. However, variations within the 2$\sigma$ experimental errors in
either helium-burning reaction rate do induce strong rms deviations
for the production factors far from the central values of those rates.

We want to issue a caution, however, about our very approximate
treatment of galactochemical evolution. Stars from different initial
metallicities contribute to the solar abundance pattern. Here we took
the approximation that the stars which contributed most are those of
about solar initial abundance, within roughly a factor of 2. Although
we did not try to obtain a precise quantification of the uncertainties
due to the form of the initial mass function (IMF), the results of our study 
were not changed in any significant way by substituting a Salpeter IMF 
for the Scalo IMF used throughout this study. Another
physics uncertainty which could affect the pre-supernova structure and 
supernova nucleosynthesis yields is the treatment of  
hydrodynamics including convection and boundary layer mixing such as 
overshoot and semi-convection. These uncertainties have been shown 
(\citealt{woo88}; \citealt{you05}) to have effects comparable to 
uncertainties in nuclear reaction rates, for instance, regarding predictions
of both carbon mass fraction and remnant mass. One more issue concerns the
poorly understood interactions of burning shells. These effects were 
discussed in \cite{rau02}, and we have pointed out above how they can 
affect nucleosynthesis for a 20 \Msun star. Such effects have also been 
confirmed in multi-dimensional calculations of pre-supernova stars 
(\citealt{mea06}). The effects of uncertainties in the calculation of mass 
loss and the possible effects of a binary companion could also be important.
It would be useful to have a numerical estimate of the implications of all
these uncertainties. However, to make something better than a guess would
involve a suite of calculations much larger than the already extensive set
we have performed. Eventually, perhaps, these effects will be sufficiently
well known to permit a reliable estimate of overall uncertainties. However, 
even then the large effects of uncertainties in the nuclear reaction rates 
will likely remain.

Within the scope of our study, uncertainties within the current errors 
in the rates of the helium burning reactions, both triple-$\alpha$ and
$^{12}C(\alpha,\gamma)^{16}O$ have been found to induce strong
changes in the remnant mass of massive stars, highlighting the fact 
that those rates are independently important. The changes in
remnant mass can have consequences for the typical neutron star
masses. Hence, determining the helium-burning reaction rates is an essential
ingredient to the theoretical understanding of the populations of neutron stars
and black holes.

Taken together, our results for SNII evolution support the need for
improved measurements of both the helium-burning reaction rates,
with the goal that their ratio is known to within 10\%.  This is
particularly important if predictions of average remnant masses are
to be reliable.

\acknowledgements

We thank Robert Hoffman for providing the solar abundance sets used 
in this study and Stan Woosley for helpful discussions, including
studies on the relative influence of the two reaction rates.
This research was supported in part by the US National Science
Foundation grants PHY06-06007 and PHY02-16783, the latter funding
the Joint Institute for Nuclear Astrophysics (JINA), a National
Science Foundation Physics Frontier Center. A. Heger performed his
contribution under the auspices of the National Nuclear Security
Administration of the US Department of Energy at Los Alamos
National Laboratory under contract DE-AC52-06NA25396, and has
been supported by the DOE Program for Scientific Discovery through
Advanced Computing (SciDAC; DE-FC02-01ER41176).

\end{document}